\documentclass[a4paper,11pt]{article}
\usepackage{jheppub}
\usepackage{amsfonts}
\usepackage{float}
\usepackage{amsthm}
\usepackage{slashed}
\usepackage{graphicx}
\usepackage{mathrsfs,bbm}
\usepackage{subfig}
\usepackage{booktabs}
\usepackage[numbers,sort&compress]{natbib}
\usepackage{lscape}
\usepackage{breqn}
\usepackage{amsmath,amssymb,textcomp,enumerate,alltt,xspace,multirow}
\usepackage{multicol}
\usepackage{boxedminipage}
\usepackage{comment}
\usepackage{tabularx}
\usepackage{adjustbox}
\usepackage{textcomp}
\usepackage[utf8]{inputenc}

\usepackage[colorlinks=true]{hyperref}
\usepackage[dvipsnames]{xcolor}
\hypersetup{
colorlinks=true,
linkcolor=blue,
linktocpage=true,
citecolor=violet
}

\usepackage{multirow}
\allowdisplaybreaks

\title{Fast evaluation of Feynman integrals for Monte~Carlo~generators
}

\newcommand{\liverpool}{Department of Mathematical Sciences, University of Liverpool, Liverpool L69 3BX, 
U.K.
}

\author[a]{Pau Petit~Ros\`as}
\author[a]{and William J. Torres~Bobadilla}
\affiliation[a]{\liverpool}

\emailAdd{paupetit@liverpool.ac.uk}
\emailAdd{torres@liverpool.ac.uk}

\abstract{
Building on the idea of numerically integrating differential equations satisfied by Feynman integrals, we propose a novel strategy for handling branch cuts within a numerical framework. We develop an integrator capable of evaluating a basis of integrals in both double and quadruple precision, achieving significantly reduced computational times compared to existing tools. We demonstrate the performance of our integrator by evaluating one- and two-loop five-point Feynman integrals with up to nine complex kinematic scales. In particular, we apply our method to the radiative return process of massive electron-positron annihilation into pions plus an energetic photon within scalar QED, for which we also build the differential equation, and extend it to the case where virtual photons acquire an auxiliary complex mass under the Generalised Vector-Meson Dominance model.  Furthermore, we validate our approach on two integral families relevant for the two-loop production of $t\bar{t}+\text{jet}$. The integrator achieves, in double precision, execution times of the order of milliseconds for one-loop topologies and hundreds of milliseconds for the two-loop families, enabling for on-the-fly computation of Feynman integrals in Monte Carlo generators and a more efficient generation of grids for the topologies with prohibitive computational costs.

}

\begin{document}
\maketitle

\section{Introduction}
\label{sec:intro}
The ability to calculate theoretical predictions with high precision represents a cornerstone of modern particle physics, enabling meaningful comparisons between theoretical frameworks and experimental observations across both classical and quantum contexts. 
The fundamental building blocks of these theoretical predictions are scattering amplitudes, which serve as the mathematical backbone connecting theoretical calculations to experimentally observables, such as cross sections and decay rates.

Scattering amplitudes at higher perturbative orders require the evaluation of multi-loop Feynman integrals, a computational challenge that becomes particularly formidable beyond next-to-leading order (NLO) accuracy. 
The evaluation of these integrals represents one of the most significant bottlenecks in achieving next-to-next-to-leading order (NNLO) 
theoretical predictions for processes relevant to collider physics~\cite{Heinrich:2020ybq,TorresBobadilla:2020ekr,Huss:2025nlt,Aliberti:2025beg}. 
This difficulty is especially pronounced when considering scattering processes with a high multiplicity in the external states
and involving internal massive states, where current techniques reach their limits.

The standard approach to address these computational challenges lies in the systematic application of integration-by-parts identities (IBPs)~\cite{Chetyrkin:1981qh, Laporta:2000dsw}, to reduce Feynman integrals to a basis of master integrals. 
These satisfy systems of differential equations~\cite{Kotikov:1990kg,Gehrmann:1999as}, which, when free of poles in the dimensional regulator, provide a systematic evaluation.
The canonical form of differential equations~\cite{Henn:2013pwa}, characterised by logarithmic integration kernels and $\epsilon$-factorised structure, 
has become the preferred framework for solving these integrals. For high dimensional kinematic cases, however, it is not trivial to solve the canonical differential equation once it has been built. Two of the methods for tackling this, proposed and explored extensively in the literature, are finding a solution through power series expansions~\cite{Lee:2017qql, Moriello:2019yhu} and a fully numerical integration~\cite{Caffo:2002ch}.

Recently, the {\sc Mathematica} packages {\tt DiffExp}~\cite{Hidding:2020ytt} and {\tt SeaSyde}~\cite{Armadillo:2022ugh}
have automated the first approach. While {\tt DiffExp} handles the evaluation of differential equations with real kinematics, 
{\tt SeaSyde} extends this to complex variables as well. 
Both are suitable for solving differential equations that are cast in canonical form or do not contain poles in the dimensional regulator $\epsilon$. 
The auxiliary mass flow method, which solves differential equations that might also contain poles in the dimensional regulator, 
has represented another strategy to numerically evaluate Feynman integrals~\cite{Liu:2017jxz}. 
This approach has been successfully automated in
the {\sc Mathematica} package {\tt AMFLow}~\cite{Liu:2022chg}. While these {\sc Mathematica}-based implementations have proven highly effective for specific calculations and boundary condition determination, 
their computational architecture presents limitations for integration into high-performance Monte Carlo event generators. Notable developments have begun to address these performance limitations through lower-level implementations, 
most notably the {\tt Line} package~\cite{Prisco:2025wqs}, which provides a {\tt C} implementation based on the auxiliary mass flow method and numerically solves differential equations via series expansion across different regions. However, current limitations in the numerical evaluation of algebraic functions appearing in differential equations continue to constrain the full realisation of these computational approaches.

The second method to solve the differential equations, that is, a fully numerical integration, was first applied to a one-dimensional physical problem in~\cite{Boughezal:2007ny}, and the use of analytical continuations to avoid singular points was proposed in~\cite{Czakon:2008zk}. Recently,  numerical integration has been used for two-dimensional processes, and better numerical algorithms have also been explored~\cite{Mandal:2018cdj,Czakon:2020vql,Czakon:2021yub,Haisch:2024nzv}. However, the effectiveness of the method remains largely unexplored for five-point scattering processes, cases in which interpolating precomputed grids becomes challenging due to the higher dimensionality of the kinematic configuration, and thus renders the power series expansion method difficult to use. With a fully numerical integration, if sufficiently low run-time could be achieved, one could bypass the need of precomputed grids and rely instead on on-the-fly evaluation of Feynman integrals. The difficulties in navigating the complex plane and the lack of public and flexible tools have prevented the use of this method, giving preference to power series expansion solutions. In this work, we bridge this gap by providing a comprehensive {\tt C++} framework for the fast numerical evaluation of systems of differential equations.

~

Our implementation is designed with the objective of direct interfacing with Monte Carlo event generators~\cite{Campbell:2022qmc,Aliberti:2024fpq}, 
eliminating the need for pre-computed grids
that become impractical when the number of kinematic scales exceeds manageable limits. Furthermore, the integrator can also be used to generate grids when high precision is needed, being easy to parallelise and achieving significantly lower run-times when compared with other tools. 
A key feature of our procedure is the treatment of algebraic functions in differential equations. 
We detail how their analytic structure can be studied and continued across physical regions, 
enabling robust and stable numerical integration. 

The structure of our implementation is motivated by the organisation of scattering amplitudes into a set of independent transcendental functions. Following the framework developed in~\cite{Chicherin:2021dyp,Gehrmann:2024tds}, we express the scattering amplitudes in terms of graded transcendental functions,  simplifying their analytic structure, and numerically evaluating differential equations for these functions~\cite{Caron-Huot:2014lda}.
The interplay between Feynman integrals and Feynman integrands is fully exploited to reduce computing time.

We demonstrate our methodology through the application to the radiative return process in electron-positron annihilation into hadrons, specifically focusing on the production of charged-pion pairs accompanied by an energetic photon.
We concentrate on the one-loop contribution to $e^+e^-\to\pi^+\pi^-\gamma$ focusing on the gauge-invariant subset of diagrams involving five-point one-loop integrals with full dependence on lepton and pion masses---corresponding to a process with seven kinematic scales. We compute these integrals to finite order in $\epsilon$ (relevant to NLO accuracy) and extend them to higher orders to prepare for NNLO theoretical predictions. 
To improve the modelling of hadronic interactions, we study the Generalised Vector-Meson Dominance model (GVMD) as suggested in~\cite{Ignatov:2022iou,Colangelo:2022lzg}. This theoretical framework, which extends beyond simple scalar QED approaches to include the dynamics of vector-meson exchange, introduces additional complexity through the presence of complex masses in internal propagators. We explicitly show that our numerical framework handles these complex kinematic scales robustly.
We systematically compare the reliability and accuracy of our computational approach against benchmark computations performed
by the {\tt Collier}  library~\cite{Denner:2016kdg,Denner:2002ii,Denner:2005nn,Denner:2010tr} for one-loop processes. 
We validate our results in both standard kinematic points and challenging regions of phase space where conventional methods may encounter numerical instabilities, elucidating the utility of our approach for phenomenological applications requiring high-precision theoretical predictions.
Beyond low-energy physical processes, we explore the applicability of our procedure to high-energy processes, 
such as the annihilation of two partons into a heavy-quark pair and an energetic jet, 
which share structural similarities with the radiative return process at low energies.
We study the performance of numerical evaluation of selected two-loop five-point Feynman integrals
with six kinematic scales within our procedure. 
We verify the reference points given in~\cite{Badger:2024fgb}.

This paper is organised as follows. In Section \ref{sec:DE}, we review the theory on differential equations for Feynman integrals, as well as how to construct a basis of graded functions. We then proceed to explain the intricacies of the numerical integrator in Section \ref{sec:NumIntegration}, with each subsection dedicated to a key part of it. Following this, we give details of the differential equation built for the $e^+e^- \to \pi^+\pi^-\gamma$ process, extended to include GVMD, and use the integrator to solve it in Section \ref{sec:Results}. In the same section, we also explore using the integrator to evaluate the amplitude of the pion process at higher orders in the dimensional regulator, as well as integrating two-loop topologies relevant for $pp\to t\bar{t}+\text{jet}$. We suggest avenues for improvement and give our conclusions in the two final sections.

~

The supplemental material of this paper is provided at~\cite{petit_rosas_2025_15920324}.
We include files containing the differential equations obtained for the one-loop process $e^+e^-\to\pi^+\pi^-\gamma$ for the three integral families, \{MM,MN,NN\}, described in Section~\ref{sec:ResultsMM}. 
For each topology, we include the definitions of the integral family, the set of canonical master integrals, the square roots associated with the basis, the kinematic alphabet, and the connection matrices. 
In addition, we provide the one-loop contribution to the pentabox gauge invariant group, as described in Section~\ref{sec:ResultsNN}, up to $O(\epsilon^2)$.

\section{Differential equations for Feynman integrals}\label{sec:DE}

A common form for a scalar $l$-loop Feynman integral is
\begin{equation}\label{eq:feynint}
    I^{d_0}_{a_1,\dots,a_N} = \int \left( \prod_{j=1}^{l} \mathrm{d}^{d_0-2\epsilon} k_j \right) \frac{1}{\prod_{i=1}^{N} D_i^{a_i}}\,,
\end{equation}
where $D_i$ are the propagators, which encode the external kinematics and the masses of the particles, $k_j$ are the loop momenta, $N$ denotes the number of propagators, and each of them is elevated to an integer power $a_i$; $\epsilon$ is the dimensional regulator, which together with $d_0$ encode the dimensions in which momenta live in. Integrals that share a set of propagators are said to be of the same family, and can be related to each other through IBPs. Furthermore, any topology can be expressed in terms of a subset of linearly independent integrals of the same family, which generate a basis. The choice of basis is not unique, and the spanning integrals are referred to as master integrals. 

It is well known that, when considering a vector $\vec{I} = (I_1,...,I_n)$ of $n$ Feynman integrals that span a basis, we can build a closed system of coupled differential equations. By taking a partial derivative of the vector with respect to a kinematic invariant $x$ and then rewriting the resulting integrals back to the same basis, through IBPs, we can obtain
\begin{equation} \label{eq:gendif}
    \partial_x \vec{I} = \mathbb{A}(\vec{x}, \epsilon) \vec{I}\,,
\end{equation}
where $\mathbb{A}$ is a matrix with rational functions that depends on kinematic invariants and $\epsilon$, the dimensional regulator. Furthermore, as extensively used in the literature and first explored in~\cite{Henn:2013pwa}, one can sometimes factor out the dimensional regulator from $\mathbb{A}(\vec{x},\epsilon)$, by studying in detail the analytic properties of the relevant Feynman integrals. 
For instance, the $\epsilon$-factorised form of one-loop topologies is easily obtained by studying their leading singularities, which are connected to Landau singularities. Although there is no general procedure to find this factorisation for integrals with higher loops, it has been found for many relevant processes, allowing us to rewrite Equation \eqref{eq:gendif} to be polynomial in the dimensional regulator,
\begin{equation}
\label{eq:polydifeq}
 d \vec{J} = \sum^\infty_{i=0}\epsilon^i\mathbb{A}^{(i)}(\vec{x})\vec{J}\,,
\end{equation}
where $\vec{J} = \mathbb{B}(\vec{x},\epsilon)\vec{I}$,
and $\mathbb{B}$ is a matrix whose entries can exhibit algebraic dependence on the kinematic variables $\vec{x}$ and the dimensional regulator $\epsilon$.
With an initial condition, the equation becomes an initial value problem (IVP) with general solution
\begin{equation} \label{eq:pathordered}
    \vec{J} = \mathcal{P} \left\{\exp\left(\int_\gamma\sum^\infty_{i=0}\epsilon^i\mathbb{A}^{(i)}(\vec{x})\right)\right\}\vec{J}(\gamma(0))\,,
\end{equation}
where $\mathcal{P}$ is the path-ordering operator and $\gamma$ is a path in the complex phase-space of $x$, connecting initial and final point. As extensively explored in the literature, analytical solutions can be found in terms of Chen iterated integrals~\cite{Chen:1977oja}, or the integrals can be evaluated along paths by series expansions \cite{Hidding:2020ytt, Armadillo:2022ugh}. Nevertheless, as explored in this work, Eq. \eqref{eq:polydifeq} can also be solved fully numerically, enabling a fast evaluation of the integrals with sufficient precision.

A further simplification of the problem might be possible by casting the differential equation in canonical form. In essence, we want to express the relevant basis of master integrals in the compact form,
\begin{equation}\label{eq:canonical}
 d\vec{J} = \epsilon \sum^n_{i=1} \mathcal{A}_i\, d\log(\alpha_i(\vec{x}))\, \vec{J}\,,
\end{equation}
where now not only does the dimensional regulator factorise in front of the sum, but also $\mathcal{A}_i$ is a matrix with only numerical coefficients, and $\alpha_i$ are what is known in the literature as the letters. The set of all letters is called the alphabet. We expect the alphabet to be composed of rational letters, coming from the denominators of the differential equations, and algebraic letters, which will depend on polynomials of the kinematic variables $P(\vec{x}), Q(\vec{x})$ and up to two of the square roots of the system, $r_i$. These letters can be cast into Galois group manifest forms, 
\begin{align}
    \alpha_i = \frac{P(\vec{x})+Q(\vec{x})r_k}{P(\vec{x})-Q(\vec{x})r_k}\,,  && \alpha_i = \frac{P(\vec{x})+Q(\vec{x})r_kr_j}{P(\vec{x})-Q(\vec{x})r_kr_j}\,.
\end{align}

It is not always possible to cast an $\epsilon$-factorised differential equation in canonical form. In these cases, we aim at finding differential equations in terms of linearly independent non-logarithmic one-forms, such that
\begin{equation} \label{eq:oneforms}
    d\vec{J} = \sum_{k=0}^\infty \epsilon^k \left[\sum_{i} \mathcal{A}_{k,i} d\log(\alpha_i(\vec{x})) + \sum_j \mathcal{B}_{k,j} \Omega_j(\vec{x})\right]\,,
\end{equation}
with $\mathcal{A}, \mathcal{B}$ being matrices of rational numbers and $\Omega(\vec{x})$ representing non-logarithmic one-forms, rational functions which might be multiplied by a square root of the system.

~

Up to this point, we have focused mainly on the evaluation of multi-loop Feynman integrals using the method of differential equations. However, our ultimate goal is to allow for numerical evaluations of scattering amplitudes that lead to theoretical predictions for physical observables. 
With the analytic techniques developed in this work, it becomes crucial to target only those quantities that directly contribute to the physical amplitude.

Following the approach proposed in~\cite{Caron-Huot:2014lda} for the analytic evaluation of finite scattering amplitudes, one can construct systems of differential equations that are independent of the dimensional regulator $\epsilon$. These systems involve only the transcendental functions relevant to the physical amplitude, thereby eliminating the need to compute redundant contributions.
More concretely, by analysing Eq.~\eqref{eq:oneforms} and accounting for the $\epsilon$ expansion of the master integrals $\vec{J}$, one can identify and isolate the differential equations that obey the physically relevant components. 
In addition to this simplification, we further organise the transcendental functions based on their analytic structure. This is achieved via their symbol map~\cite{Goncharov:2010jf}, and adopting the strategy outlined in~\cite{Chicherin:2021dyp,Gehrmann:2024tds}.
Such an organisation of the integrals in terms of linearly independent transcendental functions allows for systematic analytic cancellations when constructing the full amplitude. As a result, spurious integration kernels ---those that arise in intermediate steps but are absent in the final physical result--- are effectively removed. This streamlines the analytic structure of the amplitude and significantly improves its numerical evaluation.

To make this structure explicit, we perform a rotation of the master integrals $\vec{J}$ onto a new basis $\vec{W}$ that manifests the grading by transcendental weight,
\begin{align}
\vec{W}(\vec{x};\epsilon)&=R\,\vec{J}(\vec{x};\epsilon)\,,
\end{align}
where $R$ is a rational $\mathbb{Q}$-matrix. The new integrals $\vec{W}$ admit an expansion of the form:
\begin{align}
W_{i_{k}}\left(\vec{x};\epsilon\right) & =\sum_{k'=k}^{m}\epsilon^{k'}\,w_{i_{k'}}\left(\vec{x}\right)\,.
\end{align}
Here, the subscript $i_{k}$ labels the $i$-th canonical integral that first appears at transcendental weight $k$, following the conventions of~\cite{Bonetti:2025vfd}. As per construction, the coefficients satisfy $w_{i_{k'}} = 0$ for $k' < k$.
In Section~\ref{sec:ResultsNN}, we apply this formalism order-by-order in $\epsilon$ to the analytic evaluation of the five-point scattering process $e^+e^- \to \pi^+\pi^-\gamma$, showcasing its effectiveness in a physically relevant context.

\section{Numerical integration of differential equations} 
\label{sec:NumIntegration}

Determining what numerical algorithm to employ is key to an efficient numerical evaluation of differential equations. Following similar studies \cite{Czakon:2008zk, Mandal:2018cdj,Czakon:2020vql}, we focus on explicit methods to solve the IVP and decide to define our integrals to be complex numbers, instead of decoupling the system into real and imaginary components. The advantage of this is twofold: the overall expressions are more compact, and the vanishing imaginary parts of the integrals do not influence the error control (see Section \ref{sec:numerror}). However, this also limits the software available in the market, as most of the libraries only support real arithmetic. We decide to use the \texttt{Boost Odeint} \texttt{C++} library \cite{10.1063/1.3637934} and explore three integrating algorithms, namely \texttt{bulirsch\_stoer} (BS), \texttt{runge\_kutta\_dopri5} (DPR5) and  \texttt{runge\_kutta\_cash\_karp54} (CK45). We use {\tt AMFlow} \cite{Liu:2022chg} to obtain a boundary condition at an arbitrary phase-space point, and evolve the differential equation one variable $x$ at a time to the desired final point, following a path $\gamma(x,t_x) = (\lambda(x,t_x)_1,...\lambda_m(x,t_x))$ composed of $m$ linear segments $\lambda(x,t_x)$.  
A line parameter for each kinematic variable, $t_x$, is introduced to parametrise the segments, which only depend on $t_x$ itself and the kinematic variable $x \in \vec{x}$ that we are evolving at the time, such that
\begin{equation}
    \lambda(x,t_x) = x_0 + t_x (x_f-x_0)\,.
\end{equation}
To go from the origin $x_0$ and the desired final point $x_f$, we simply evolve $t_x$ from 0 to 1. We expand the master integrals in powers of the dimensional regulator $\epsilon$, rewriting Eq.~\eqref{eq:canonical} to solve it order by order, and parametrise it with respect to $t_x$, such that
\begin{align}
    \partial_{t_x} \vec{J}^{(k)} = \epsilon^k \textbf{A}_{t_x}\vec{J}^{(k')}\,,  && \textbf{A}_{t_x} = \sum_{i=1}^n \mathcal{A}_i \frac{\partial \log(\alpha_i(x))}{\partial x}\frac{\partial x}{\partial t_x}\,.
\end{align} 
The same procedure is followed in the case where differential equations contain non-logarithmic one-forms. 

\subsection{Treatment of branch cuts} \label{sec:branchCuts}
To give an accurate evaluation of the integrals, the path should be chosen to avoid both singularities, present in the differential equation, and branch cuts, introduced by the multi-valued square roots $r_i$ present in the letters. Avoiding singularities is trivial, as we can choose our path to extend to the complex plane by means of an analytic continuation. However, the square roots that we have introduced in the master integrals present non-trivial branch cuts that, without proper treatment, could prevent us from reaching certain regions of the phase-space. Ideally, we would like to define the square roots as continuous and single-valued, but that is, of course, impossible in the entire complex plane. 

We need them to meet these conditions along the integration path $\gamma(\vec{x},t)$. If we naively proceed by writing the square roots as
\begin{align}
    r_i = \sqrt{K^n_i(\vec{x})}\,, &&  K^n(\vec{x}) = \Pi^n_{j=1}\left( x-e_j \right)\,,
\end{align}
where $K^n_x(\vec{x})$ are $n$ degree polynomials of the kinematic variables $\vec{x}$ and, for a particular variable $x \in \vec{x}$,
we factor each polynomial in terms of its roots $e_j$, the square roots will have branch cuts determined by the function \cite{Bogner:2017vim}
\begin{equation}\label{eq:brcut}
    \Pi^n_{j=1}\left( x-e_j \right) - \rho = 0\,,
\end{equation}
where $\rho$ is negative and real. These algebraic curves increase in complexity with the degree of the polynomial and can easily generate inaccessible phase-space areas, particularly if we work with complex kinematic scales or several square roots. Navigating the complex phase-space of the differential equation would require solving Eq.~\eqref{eq:brcut} to know where the branch cuts are. Nevertheless, if we work with the irreducible square roots,
\begin{equation}
    r_i = \Pi^n_{j=1} \sqrt{x-e_j}\,, 
\end{equation}
and bear in mind the standard convention for mathematical software to define the branch cuts along the negative real axis, we end up with much simpler to handle segments at
\begin{align}
    e_j+\rho_j\,, && \rho_j \in \mathbb{R}_{\leq 0}\,.
\end{align}

\begin{figure}[t]
    \centering
    \includegraphics[width=0.4\textwidth]{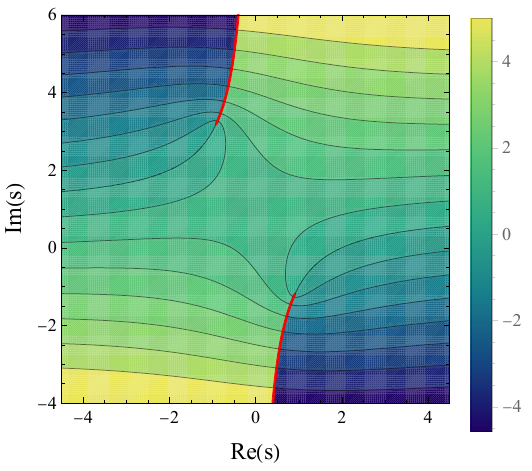}
    \includegraphics[width=0.4\textwidth]{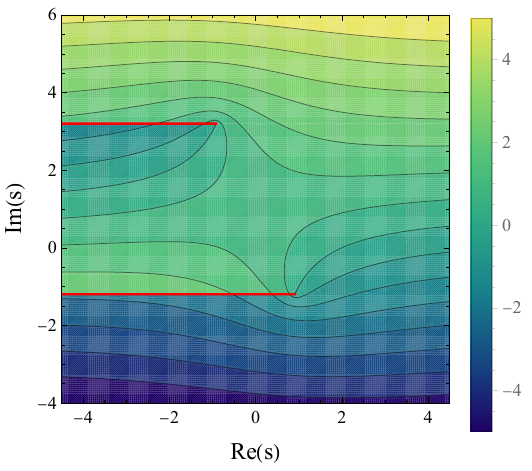}
    \caption{Value of the imaginary part of $\sqrt{\lambda_K(s,t,m)}$ written as $r_i~=~\sqrt{K^n_i(\vec{x})}$ (left) and $r_i~=~\Pi^n_{j=1} \sqrt{x-e_j}$ (right) in the complex $s$ plane. The branch cuts are depicted with light red lines.}
    \label{fig:branchcut}
\end{figure}

We plot an example of the result of this manipulation in Fig.~\ref{fig:branchcut}, where we represent the value of a single square root $\sqrt{\lambda_K(s,t,m)}$,\footnote{Here $\lambda_K$ stands for the Källén function $\lambda_K(a,b,c)~=~a^2+b^2+c^2-2ab-2ac-2bc$} in the complex plane of $s$, with arbitrary complex values $m = 1+i$ and $t = -1$. Although the square root only contains a polynomial of degree 2 in $s$, it is easy to see that a differential equation with several of these might have isolated regions in the complex plane. With this definition of the square roots, we ensure that we can find a single-valued path that connects two points in the phase-space.

The reader may have noticed that the square roots change definition depending on the kinematic variable $x$ we choose. This is indeed the case, but we are free to redefine them arbitrarily, as long as they keep our master integrals as functions with uniform degree of transcendentality. Since the integration is performed variable by variable, we define the square roots taking $x$ as the variable that is currently being integrated. Once the integration for $x$ is complete, we can divide the result by the Landau singularity factors. If we then need to do another integration, we can add them back by multiplying the square root free result with the factors defined for the new variable.

\subsection{Finding an integration path}
\label{sec:path}
We have outlined a way to treat branch cuts in differential equations that simplifies their behaviour. Nevertheless, we still need to find an optimal path that avoids crossing them, together with the singularities, and that connects initial and final points. This is highly non-trivial. We make two assumptions to simplify the problem: we rely on the heuristic of Euclidean distance between points and, as mentioned at the beginning of this section, work with linear segments. These are strong assumptions; a more complex path could avoid regions of large variations, and a different set of kinematic variables could be more optimal. However, our approach is sufficient for the examples at hand, and we briefly discuss other possibilities in Section~\ref{sec:futurework}.  Currently, the integrator supports two algorithms, with both first checking whether a straight path connecting the initial and final points is possible. If not, the first approach makes use of an A* algorithm \cite{Zeng01042009} to find the shortest path. In essence, we build a graph with candidate points, which are our boundary value, the desired end point and a sample of points near singularities, and we use Euclidean distance as the heuristic for the algorithm to find the shortest path between nodes of the graph.
Although the approach is general and ensures a path is found between initial and final points, it forces the path to go near singularities, which slows the integrator down and worsens its precision. 

The second approach proves to be easier to implement and more effective. In practice, we check whether the initial and final points have the same imaginary part. If they do, and there are singularities in between, the path is chosen to be a horseshoe with imaginary part between the lowest branch cuts above the real axis and the real axis itself, similar to what is used in~\cite{Armadillo:2022ugh}. 
In contrast, if the initial and final points lie in a different part on the imaginary axis, the auxiliary points will be chosen so that the path encircles, again in horseshoe form, any branch cut between the initial and final point. The choice of a horseshoe path is not arbitrary. Even if the distance in the complex plane is larger than, for example, a triangular path, the CPU time is reduced from evolving only the real or the imaginary components at a time. 

There is an extra edge case that has been identified as a source of imprecision and higher run times. If there are branch cuts with a very small imaginary part that bounds the path, the analytic continuation is less effective, since we are forced to go near singularities in the real axis.  In these cases, we rotate the branch cut so that it is parallel to the negative imaginary axis, and choose a path that encircles its branch point.

\subsection{Reducing the CPU time} \label{sec:reducetime}
To further reduce computational time, we exploit the structure of the differential equation. At each step of the numerical integration, the code evaluates the analytic expression for $d\log(\alpha_i)$, one-forms, and the square roots of the system. This is then used to evaluate the right-hand side of the differential equations without the need to re-evaluate repeated analytical structures in the same integration step. We further exploit the structure of the $d\log$ and one-forms: they are rational functions (up to factors of square roots) of the form:
\begin{equation}
    \frac{P^n(\vec{y},x)}{Q^m(\vec{y},x)} = \frac{a_0(\vec{y})+a_1(\vec{y})x+\hdots + a_n(\vec{y})x^n}{b_0(\vec{y})+b_1(\vec{y})x+\hdots + b_m(\vec{y})x^m}\,,
\end{equation}
where $\vec{y}$ is the set of kinematic variables without $x$. Since we only vary one $x\in\vec{x}$ at a time, we can compute and store the coefficients $a, b$ of the $m$ and $n$ degree polynomials before evolving $x$. These can be reused for all integration steps, and their evaluation and compilation are enhanced with the {\tt FORM} optimiser \cite{Kuipers:2013pba}.\footnote{It is known that the optimiser can magnify numerical errors. We do not explore the effects of using it, but the compilation times are greatly reduced by its use. We also observe a speed-up in the CPU times.} A similar approach is used to evaluate the square roots at every step. The roots of the polynomial inside them only depend on non-evolving variables, and are only calculated once before each integration. The same is applied to the square roots that do not depend on the variable that is currently being integrated, whose value is computed once and saved.

\subsection{Error control and numerical stability}
\label{sec:numerror}

The numerical value for the master integrals is to be used in evaluating scattering amplitudes, which can amplify numerical errors, generate catastrophic cancellations, and lead to instabilities. It is necessary to determine the precision needs case by case, and a good amount of checks to make sure these are met. There are two main contributions to the final error of the integration.

The first contribution we consider is the error that comes from the numerical algorithm itself, $E$. For this reason, we use absolute and relative tolerances, $\mathcal{T}_A$ and $\mathcal{T_R}$, with values defined by the user. The implemented controlled stepper in the \texttt{Boost Odeint} library ensures that at each step the error of the value $y$ satisfies
\begin{equation} \label{eq:errtol}
    E < \max(\mathcal{T}_a, \mathcal{T}_r |y|)\,. 
\end{equation}
A good rule of thumb for determining the final relative error is to multiply the number of steps by the relative tolerance. This error is approximately linearly dependent on the total number of steps, and thus also varies with the number of variables to integrate. A caveat of defining the integrals as complex types, as hinted in the beginning of this section, is that Eq. \eqref{eq:errtol} is based on the magnitude of the complex numbers, not on their individual parts. This is helpful when considering that the integrator will be able to use the relative tolerance even when dealing with small imaginary or real parts, as long as their real or imaginary counterparts are big enough. However, the error will be minimised for the overall magnitude, not for the relevant real part of the integrals. Following the same line of thought as in \cite{Czakon:2008zk}, we conclude that this does not pose a big problem for our use case but could be something to improve in the future.

Perhaps even more relevant is the second contribution to the total numerical error, namely the numerical instabilities that arise in the differential equation itself. In canonical form, rational letters and square roots generate spurious and physical singularities when they vanish. More generally, the singularities can be determined analytically by looking at the denominators in the differential equation. The spurious poles are an artifact of the differential equation itself and are not singularities of their solution once the initial conditions are taken into account. However, both the spurious and the physical singularities are an issue for numerical methods when the integration path gets close to them, with the distance where they start posing a problem depending on both the available precision type and the strength of the singularity, determined by the power of the denominator. Using the method described in Section \ref{sec:path}, the distance to singularities should be enough to not significantly affect numerical integration. There is an important exception to this, which is in the case when the final desired point itself is near any pole. In these cases, there are several solutions that one could try. For example, for spurious singularities, the solutions could be obtained by interpolation, but this problem is already greatly mitigated by using graded functions, which analytically remove them. Higher-precision types, or alternative solutions, might be needed for final points near physical singularities. We do not explore other options in the current work and restrain ourselves to work in double or quadruple precision, enough for the examples at hand. The quadruple precision is implemented with \texttt{GCC}'s quadruple precision library, and relies on the data type \texttt{\_\_float128}. 

\subsection{A pedestrian example}
We consider a simple massive bubble Feynman integral to illustrate the process followed by the integrator. As a preliminary step, we require the differential equation, which we obtain for the basis of bubbles and tadpoles $\vec{J}=\epsilon\{ R(I_{1,2}^4+I_{2,1}^4), I_{2,0}^4, I_{0,2}^4\}$, with 
\begin{align}
    D_1 = (k^2-m^2+\imath 0)\,, && D_2 = ((k-p)^2-M^2+\imath 0)\,, && R=\sqrt{\lambda_K(s,m^2,M^2)}\,,
\end{align}
where we work with the Mandelstam variable $s = p^2$. In this case, the system of differential equations contains four rational letters and two algebraic ones, with a single square root $R$, and can be cast in the form of Eq.~\eqref{eq:canonical},
\begin{equation}
d\vec{J} = \epsilon
\begin{pmatrix}
(\beta_3-\beta_4) & \beta_6 & -(\beta_5+\beta_6) \\
0 & \beta_1 & 0\\
0 & 0 & \beta_2
\end{pmatrix}\vec{J}\,,
\end{equation}
where the six letters are defined by
\begin{align}
    \beta_i = d\log(\alpha_i), && \vec{\alpha} = \left\{m^2,  M^2, s, R^2, \frac{m^2-s+M^2-R}{m^2-s+M^2+R},\frac{m^2+s-M^2-R}{m^2+s-M^2+R}\right\}\,. 
\end{align}
We now perform the analytic expansion of the Feynman integrals in the dimensional regulator $\epsilon$, and collect the coefficients of each order in $\epsilon$ to solve the system of differential equations.

Let us start by integrating $s$, from an initial condition obtained at an arbitrary non-singular point. According to the definition of square roots proposed in Section \ref{sec:branchCuts}, $R$ will now take the form
\begin{equation}
    R = R_s = \sqrt{s-(m-M)^2}\sqrt{s-(m+M)^2}\,,
\end{equation}
from which we can easily see there will be two branch cuts, horizontal lines parallel to the negative real axis, with branch points in $(m-M)^2$ and $(m+M)^2$. We also determine the singularities of the system by looking at the points where a letter vanishes. There are seven, one coming from $\alpha_3$ ($s=0$), and the rest coming from $\alpha_4, \alpha_5$ and $\alpha_6$. With this information, it is easy to choose a path that avoids them. 

We can now parametrise each $\beta$ with respect to the variable we are currently integrating without loss of generality, in this case $s$, and save CPU time by computing only the necessary components at each integration step. Following the previous section, we calculate and store the branch points and reuse them every time we need to compute $R$. Similarly, we rewrite all $\beta$'s in rational and algebraic form. For example, the 4-{\it th} $d\log$ now reads
\begin{equation}
    \beta_4 = \frac{-2(m^2-s+M^2)}{m^2+(M^2 - s)^2 - 2 m^2 (M^2 + s)} = \frac{a_0 + a_1s}{b_0+b_1s+b_2s^2}\,. 
\end{equation}
With this, we can now proceed with numerical integration along the chosen path, without the need to reevaluate the constant coefficients $a_i$ and $b_i$, and taking into account the change of variables from $s$ to the variable of the auxiliary path, $t_{s}$. Beforehand, we also need to specify the desired values for the absolute and relative tolerances, as well as the numerical integration algorithm. 

We now desire to evolve another variable, say $M$. Following the same steps as before, we need to redefine the square root to be
\begin{equation}
    R = R_M = \sqrt{M^2-\left(\sqrt{s}-m\right)^2}\sqrt{M^2-(\sqrt{s}+m)^2}\,.
\end{equation}
However, what now will be the boundary value for $J_1^2$ has $R = R_s$, the definition of the square root that we used in the previous integration. To account for this, we divide it by $R_s$ and multiply by $R_M$. We can then proceed with the same steps that we followed in $s$ integration, redefining the $\beta$'s to be polynomials in $M$, instead of $s$.

\section{Results} \label{sec:Results}

We test the precision, speed, and usefulness of the fully numerical approach by using the integrator in three examples. 
All tests have been performed on a single thread of an \texttt{Intel(R)Core(TM) i7-13700H CPU @ 2.40 GHz}.
First, we employ the integrator to find values for five-point one-loop integrals with up to 9 complex scales for two orders in the dimensional regulator. We compare the results with~\texttt{Collier} \cite{Denner:2016kdg,Denner:2002ii,Denner:2005nn,Denner:2010tr} and demonstrate an additional use of the integrator; evolving only a subset of the variables. In the second example, we push the integrator to evaluate graded functions that appear in one-loop five-point Feynman integrals with 7 kinematic scales up to transcendental weight 4, 
in particular, for the the pentabox gauge invariant group of $e^+e^- \to \pi^+\pi^-\gamma$ up to
$\mathcal{O}(\epsilon^2)$. 
Finally, we implement two families, PB$_A$ and PB$_B$ of the two-loop process $p\bar{p} \to t\bar{t}+$jet, from~\cite{Badger:2024fgb}, and evolve up to transcendental weight four from an initial point to a benchmark value, both provided in the same reference. We comment on the precision and speed reached by the integrator, suggest avenues for improvement, and give an estimate of the error. 

\subsection{One-loop five-point integrals with up to 9 complex kinematic scales} \label{sec:ResultsMM}

\begin{figure}[t]
    \centering
    \includegraphics[width=0.5\linewidth]{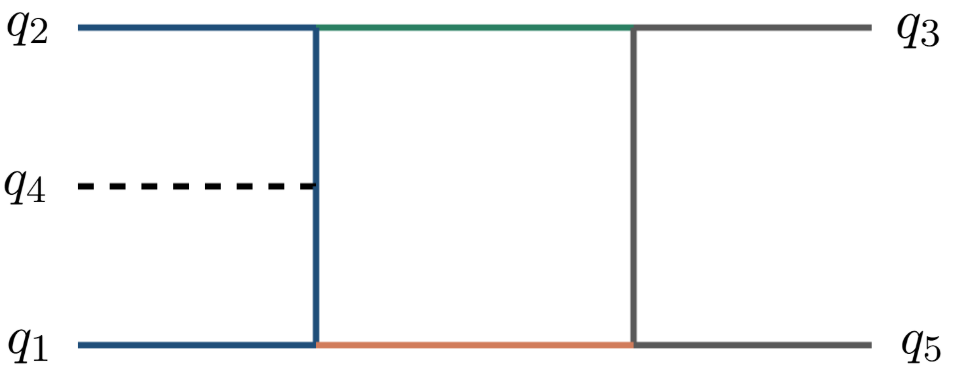}
    \caption{Pentagon topology studied in the first example. Dashed line denotes a massless particle, while solid lines of different colours carry different masses. The green and orange internal propagators can be massless or have a complex mass. See the text for more details. 
    }
    \label{fig:penttopo}
\end{figure}

We test the functionality of our integrator in the five-point scattering process, 
\begin{subequations}
\begin{align}
q_1+q_2+q_3+q_4+q_5 = 0\,,
\end{align}
with the kinematic invariants, 
\begin{align}
 & s_{14}=\left(q_{1}+q_{4}\right)^{2}\,, &  & s_{23}=\left(q_{2}+q_{3}\right)^{2}\,, &  & s_{15}=\left(q_{1}+q_{5}\right)^{2}\,,\nonumber \\
 & s_{24}=\left(q_{2}+q_{4}\right)^{2}\,, &  & s_{35}=\left(q_{3}+q_{5}\right)^{2}\,,
\end{align}
and massive and massless external momenta, 
\begin{align}
 & q_{1}^{2}=q_{2}^{2}=m_{1}^{2}\,, &  & q_{3}^{2}=q_{5}^{2}=m_{2}^{2}\,, &  & q_{4}^{2}=0\,.
\end{align}
\label{eq:5pt_kin}
\end{subequations}
As a first example, we implement a one-loop five-point topology with 7, 8 and 9 complex kinematic scales, displayed in Fig.~\ref{fig:penttopo}. All internal lines can be massive, with masses from the external momenta, $m_1$ and $m_2$, or auxiliary complex masses $\Lambda_w$ and $\Lambda_v$. This topology appears in the process $e^+e^- \to \pi^+\pi^-\gamma$ described by the GVMD model~\cite{Ignatov:2022iou}, whose preliminary implementations do not include the final-state photon~\cite{Arbuzov:2005pt, Budassi:2024whw,Colangelo:2022lzg}. 
In this case, a fast numerical evaluation of these topologies is key because the integrals are needed with different values of the auxiliary masses. 
In total, each integral needs to be evaluated 16 different times, with either the auxiliary masses set to zero or a combination of 3 different values. 

We consider three integral families, namely, the family with the two auxiliary masses set to zero (NN), 
the two auxiliary masses being massive (MM),
and only one being massless (MN). These integral families are listed in
Table~\ref{tab:props} for the three groups \{MM, MN, NN\}. Furthermore, we also report features of the canonical differential equations obeyed by the set of integrals appearing in the above families. We build the differential equation following the literature.

First, we construct our canonical basis such that each element is a pure function of uniform transcendental weight. 
This is achieved by analysing integrals in various space-time dimensions and identifying their corresponding leading singularities, following~\cite{Flieger:2022xyq}. In particular, we consider tadpole and bubble topologies in $D = 2 - 2\epsilon$, triangle and box integrals in $D = 4 - 2\epsilon$, and pentagon integrals in $D = 6 - 2\epsilon$. To construct the differential equations, we make use of the dimensional recurrence relations~\cite{Lee:2009dh} implemented in {\tt LiteRed}~\cite{Lee:2012cn,Lee:2013mka}, which allow us to express all integrals consistently in $D = 4 - 2\epsilon$.

 Secondly, we predict the letters of the kinematic alphabet that appear in our differential equations with dedicated programs \texttt{BaikovLetters}~\cite{Jiang:2024eaj} and \texttt{Effortless}~\cite{Antonela}. 
 Finally, we use the letters as an ansatz to reconstruct Eq.~\eqref{eq:canonical} over finite fields using {\tt FiniteFlow}~\cite{Peraro:2019svx}. 
 Let us remark that for the two topologies with at least one $\Lambda$ being non-zero, the amount of odd letters surpasses the number of even ones, which shows the complexity of the system.  

\begin{table}[t]
    \centering
    \renewcommand{\arraystretch}{1.3}
        \begin{tabular}{>{\centering\arraybackslash}p{2cm}  >{\centering\arraybackslash}p{3cm} >{\centering\arraybackslash}p{3cm} >{\centering\arraybackslash}p{3cm} >{\centering\arraybackslash}p{3cm}}
        \toprule
            & MM & MN & NN \\
            \midrule
            $D_1$ & $\Lambda_w^2 - p^2$ & $- p^2$ & $- p^2$ \\
            $D_2$ & $m_2^2 - (p-q_3)^2$ & $m_2^2 - (p-q_3)^2$ & $m_2^2 - (p-q_3)^2$ \\
            $D_3$ & $m_1^2 - (p+q_2)^2$ & $m_1^2 - (p+q_2)^2$ & $m_1^2 - (p+q_2)^2$ \\
            $D_4$ & $\Lambda_v^2 - (p+q_{124})^2$ & $\Lambda_v^2 - (p+q_{124})^2$ & $- (p+q_{124})^2$ \\
            $D_5$ & $m_1^2 - (p+q_{24})^2$ & $m_1^2 - (p+q_{24})^2$ & $m_1^2 - (p+q_{24})^2$  \\
            \midrule
            \# $\alpha_{even}$ & 42 & 44 & 43 \\
            \# $\alpha$ & 142 & 104 & 67 \\
            \# $r_i$ & 20 & 14 & 9\\
            \# MIs & 29 & 25 & 21 \\
            \bottomrule
        \end{tabular}
    \caption{Propagators, number of even and total letters $\alpha$, number of square roots $r_i$ and number of master integrals (MIs) for the three different topologies. Note that $q_{ij}=q_i+q_j$ and $q_{ijk}=q_i+q_j+q_k$.}
    \label{tab:props}
\end{table}

~

We implement the three differential equations in our integrator, working with the 9 kinematic complex variables\footnote{Only the auxiliary masses $\Lambda_v$ and $\Lambda_w$ are evolved to a final value with an imaginary part. However, we work with all the kinematic variables being complex numbers, a requirement for the analytic continuation.}
\begin{equation}
\vec{x} = \{s_{14},s_{15},s_{23},s_{24},s_{35},m_1^2,m_2^2,\Lambda_v^2,\Lambda_w^2\}\,,
\end{equation}
as defined in Eqs.~\eqref{eq:5pt_kin}, and proceed to check the results for the MM topology.  For a robust validation of the numerical precision, we compare the integrals with \texttt{Collier}, with which we obtain the value of the master integrals for $10^5$ phase-space points. The basis is defined as
\begin{align}
    J_1  &= \epsilon^3 r_1 I^{6;\text{MM}}_{1,1,1,1,1}, \notag \\
    J_2  &= \epsilon^2 r_2 I^{4;\text{MM}}_{1,0,1,1,1},  &  
    J_3  &= \epsilon^2 r_3 I^{4;\text{MM}}_{1,1,0,1,1},  &
    J_4  &= \epsilon^2 r_4 I^{4;\text{MM}}_{1,1,1,1,0},  \notag \\ 
    J_5  &= \epsilon^2 r_5 I^{4;\text{MM}}_{1,1,1,0,1},  &  
    J_6  &= \epsilon^2 r_6 I^{4;\text{MM}}_{0,1,1,1,1},  &
    J_7  &= \epsilon^2 r_7 I^{4;\text{MM}}_{1,0,0,1,1},  \notag \\  
    J_8  &= \epsilon^2 r_8 I^{4;\text{MM}}_{1,0,1,1,0},  &  
    J_9  &= \epsilon^2 r_9 I^{4;\text{MM}}_{1,1,0,1,0},  &
    J_{10} &= \epsilon^2 \left(m_1^2 - s_{24}\right) I^{4;\text{MM}}_{1,0,1,0,1}, \notag \\  
    J_{11} &= \epsilon^2 r_{10} I^{4;\text{MM}}_{1,1,0,0,1}, &  
    J_{12} &= \epsilon^2 r_{11} I^{4;\text{MM}}_{1,1,1,0,0}, &
    J_{13} &= \epsilon^2 \left(m_1^2 - s_{14}\right) I^{4;\text{MM}}_{0,0,1,1,1}, \notag \\  
    J_{14} &= \epsilon^2 r_{12} I^{4;\text{MM}}_{0,1,0,1,1}, &  
    J_{15} &= \epsilon^2 r_{13} I^{4;\text{MM}}_{0,1,1,1,0}, &
    J_{16} &= \epsilon^2 \left(s_{15} - s_{23}\right) I^{4;\text{MM}}_{0,1,1,0,1}, \notag \\
    J_{17} &= \epsilon r_{14} I^{2;\text{MM}}_{1,0,0,1,0},  &  
    J_{18} &= \epsilon r_{15} I^{2;\text{MM}}_{0,1,0,1,0},  &
    J_{19} &= \epsilon r_{12} I^{2;\text{MM}}_{0,1,0,0,1},  \notag \\  
    J_{20} &= \epsilon r_{11} I^{2;\text{MM}}_{0,1,1,0,0},  &  
    J_{21} &= \epsilon r_{16} I^{2;\text{MM}}_{1,1,0,0,0},  &
    J_{22} &= \epsilon r_{17} I^{2;\text{MM}}_{0,0,0,1,1},  \notag \\  
    J_{23} &= \epsilon r_{18} I^{2;\text{MM}}_{0,0,1,1,0},  &  
    J_{24} &= \epsilon r_{19} I^{2;\text{MM}}_{1,0,0,0,1},  &
    J_{25} &= \epsilon r_{20} I^{2;\text{MM}}_{1,0,1,0,0}, \notag \\  
    J_{26} &= \epsilon I^{4;\text{MM}}_{0,0,2,0,0},  &  
    J_{27} &= \epsilon I^{4;\text{MM}}_{0,2,0,0,0},  &
    J_{28} &= \epsilon I^{4;\text{MM}}_{0,0,0,2,0},  \notag \\  
    J_{29} &= \epsilon I^{4;\text{MM}}_{2,0,0,0,0}, & &  
\end{align}
 where the square roots $r_i$ are written in terms of Gram and Cayley determinants, as well as Källén functions, defined in the ancillary files. The additional superscript MM denotes integrals with propagators of this particular family.

 The Mandelstam variables are constrained to be between $\pm 15$, $m_1$ and $m_2$ are set to the mass of the electron and pion respectively, and $\Lambda_v = \Lambda_w= 0.57032704-0.099610879i$, a realistic value for the GVMD model obtained from~\cite{Ignatov:2022iou}.

\begin{figure}[t]
    \makebox[\textwidth][c]{
        \includegraphics[width=1.1\textwidth]{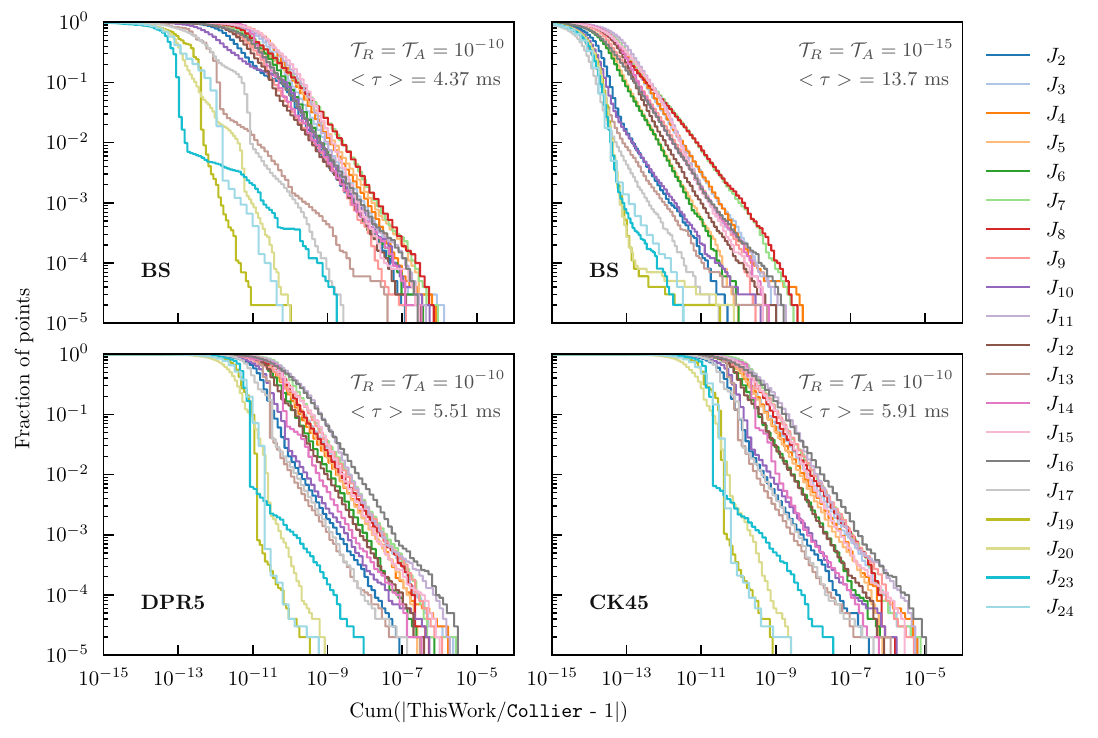}
    }
    \caption{Fraction of points versus the cumulative relative precision reached as compared with \texttt{Collier}, for 20 master integrals evaluated at $10^5$ phase-space points, integrated with BS (top row), DPR5 (bottom left) and CK45 (bottom right) algorithms. The absolute and relative tolerance parameters, $\mathcal{T}_a$ and $\mathcal{T}_r$, and the average time per phase-space point $<\tau>$ are also displayed.  }
    \label{fig:precs}
\end{figure}
 
 Since \texttt{Collier} only calculates integrals up to finite order in the dimensional regulator, the checks can only be done up to this coefficient. We present the comparison with our integrator for the master integrals of the MM topology in Fig.~\ref{fig:precs}. In it, we display the cumulative sum of the relative difference between our integrator and \texttt{Collier}, versus the fraction of the $10^5$ points, for all the real finite order coefficients of the master integrals. We dedicate one panel to each of the three algorithms supported in the integrator, with relative and absolute tolerances set to $10^{-10}$. For reference, we also plot one panel where the tolerances are reduced to $10^{-15}$.  We exclude from the plot the pentagon $J_1$, which appears at transcendental weight three, and the bubbles $J_{18}$, $J_{21}$, $J_{22}$ and $J_{25}$ as well as the tadpoles, since they would only change with a changing $m_e$, $m_\pi$ and $\Lambda_{v,w}$. Note, however, that tadpoles and bubbles also contribute to the total run times. 
 
 The three algorithms reach similar precision. For all points, the agreement is better than 5 significant figures. Differences between algorithms are mild, but the BS algorithm performs slightly better in both run-time and precision. The CPU time of the three algorithms is also similar, with BS taking an average time per phase-space point $< \tau>~=~4.37$~ms, and DPR5 and CK45 taking 5.51 and 5.91 ms, respectively. For the BS integrator, the lowest bound on the run-time
is $\mathcal{O}( 2\text{ ms})$, while a few points take up to $16$~ms. Similar numbers, albeit larger, are observed for the two other numerical algorithms. The BS algorithm seems to be the best for the integrals at hand, although by a small margin. This is not surprising; BS algorithm shines when the integrated functions are very smooth and low tolerances are required. Its superiority was also identified in \cite{Czakon:2020vql}. We already anticipate here that the differences between algorithms are bigger for more complex integrals, and in the other two examples the BS algorithm is the preferred one. 

CPU time and precision are highly dependent on absolute and relative tolerances. We show this in the upper right panel, where we repeat the evaluation using the BS algorithm but reducing the relative and absolute tolerances to $10^{-15}$. The mean CPU time per point triples, but the precision increases by more than 2 significant figures. Unsurprisingly, the bubble integrals of the system are the topologies with the best precision achieved. In all panels, a triangle is the integral with worse precision. With error requirements varying between applications, it is clear that extensive and individual checks are needed to verify that these are met. 

An advantage of evolving each kinematic variable individually is the possibility of getting to a new phase-space point at a reduced cost, when only a subset of the scales changes. This can be very useful when, as in the current example, the same integral with different masses is needed. However, and arguably more importantly, this also allows us to get the value of the master integrals with crossed external kinematics at a fraction of the run time.

For the example at hand, we evaluate the MM topology to a certain kinematic phase-space, and then evolve the auxiliary masses from $\Lambda_v = \Lambda_w= 0.57032704-0.099610879i$ to $\Lambda_v = \Lambda_w= 1.404225-0.679242i$ and finally $\Lambda_v = \Lambda_w= 2.985984-0.3274560i$, all realistic numbers obtained from~\cite{Ignatov:2022iou}, getting three different values for the master integrals. Returning to the relative and absolute tolerances set to $10^{-10}$, the integrator takes, on average, an extra $\thicksim 4$ ms to finish, effectively doubling the computation time, but giving three times the information. We expect a reduction on the precision due to the four additional integrations and, indeed, we observe an increase on the overall error. However, if one uses the BS algorithm, the agreement remains better than the one obtained in Fig.~\ref{fig:precs} using the CK45 algorithm.

\subsection{One-loop $e^+e^- \to \pi^+\pi^-\gamma$ in sQED at $O(\epsilon^2)$} \label{sec:ResultsNN}

\begin{figure}[t]
    \centering
    \includegraphics[width=\textwidth]{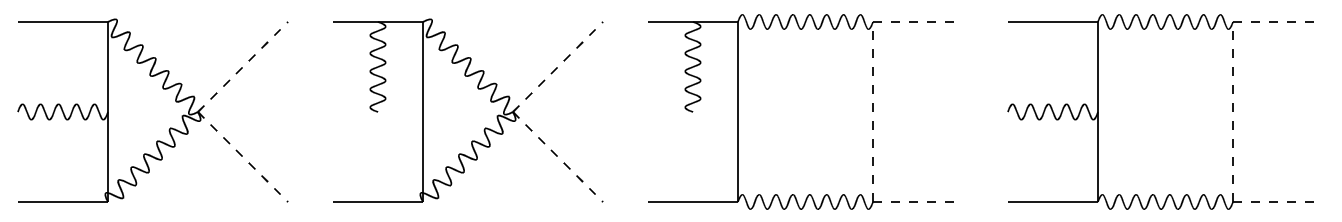}
    \caption{Representative diagrams for the pentabox gauge invariant group for the process $e^+e^-\to\pi^+\pi^-\gamma$. Note that we omit in the picture the diagrams where the photon is emitted from the other external lepton line, as well as diagrams with crossing of the external pions.
    }
    \label{fig:piondiags}
\end{figure}

An efficient evaluation of one-loop integrals at higher orders in $\epsilon$ is vital to compute full NNLO amplitudes. To show the validity and capabilities of our integrator, we calculate the
interference of the one-loop ($A^{(1)}_{\text{ISR}}$) and Born ($A^{(0)}_{\text{ISR}}$)
amplitudes of the initial state radiation (ISR) pentabox gauge invariant group, with the one-loop diagrams shown in Fig.~\ref{fig:piondiags}, for the process 
\begin{equation}
  e^-(q_1)+e^+(q_2)\xrightarrow{}\pi^-(-q_3)+\gamma(-q_4)+\pi^+(-q_5) \,, 
\end{equation}
  with full masses for electrons and pions, 
  according to~\eqref{eq:5pt_kin}. We work with the set of kinematic variables used in the previous example with $\Lambda_w= \Lambda_v=0$ and generate the relevant topologies with \texttt{qgraf}~\cite{Nogueira:1991ex}. We dress them with scalar QED Feynman rules using a dedicated model built in \texttt{Tapir}~\cite{Gerlach:2022qnc}. The Dirac algebra, integral identification, and an expansion in the dimensional regulator of the amplitude is processed by a {\tt FORM}~\cite{Vermaseren:2000nd,Ruijl:2017dtg,Kuipers:2012rf} custom script. \texttt{Kira}~\cite{Maierhofer:2017gsa,Maierhofer:2018gpa,Klappert:2020nbg} is then used to reduce all integrals to a set of master integrals, which are equivalent to topology NN in the previous example. 

To exploit the analytical structure of the final desired amplitude we build the system of graded functions considering the Feynman diagrams in Fig.~\ref{fig:piondiags} and their crossed counter part, obtained by crossing the external momenta of the pions $q_3 \leftrightarrow{} q_5$. 

The infrared structure of this one-loop amplitude simply becomes
\begin{align}
A_{\text{ISR}}^{\left(1\right)}\times A_{\text{ISR}}^{\left(0\right)\,*}&=\frac{2}{\epsilon}\left|A_{\text{ISR}}^{\left(0\right)}\right|^{2}
\left(\frac{1}{v_{15}}w_{2_{1}}+\frac{1}{v_{23}}w_{3_{1}}+\frac{1}{v_{13}}w_{4_{1}}+\frac{1}{v_{25}}w_{5_{1}}\right)
+\mathcal{O}\left(\epsilon^{0}\right)\,,
\end{align}
with $v_{ij}=\sqrt{1-\frac{4m_{1}^{4}m_{2}^{4}}{\left(s_{ij}-m_{1}^{2}-m_{2}^{2}\right)^{2}}}$. The transcendental functions obey the differential equations, 
\begin{align}
d\vec{\omega} & =d\Xi\,\vec{\omega}\,,
\end{align}
with, $\vec{\omega}=\left\{ w_{1_{0}},w_{2_{1}},w_{3_{1}},w_{4_{1}},w_{5_{1}}\right\}  \,,\Xi^{T}=\left\{ \left\{ 0,L_{15},L_{23},L_{13},L_{25}\right\} ,\vec{0},\hdots,\vec{0}\right\} $,
and the integration kernels, $L_{ij}=-\frac{1}{4}\log\left(\frac{1-v_{ij}}{1+v_{ij}}\right)$.
In this representation, it is worth noting that analytically solving the differential equations naturally reveals the infrared structure predicted by the Catani operator~\cite{Catani:2000ef}.

For the finite and higher-order contributions to the amplitude up to $\mathcal{O}(\epsilon^2)$, we find 31, 60 and 89 transcendental functions. Nevertheless, instead of using the system of differential equations with the 89 transcendental functions, we only implement it for the 65 functions that do not come from the crossing of external momenta. One could obtain values for the functions and then evolve $s_{15}\to s_{13}$ and $s_{23}\to s_{25}$ to get the crossed values. 

\begin{figure}[t] 
    \centering
\makebox[
\textwidth][c]{%
        \includegraphics[width=0.8\linewidth]{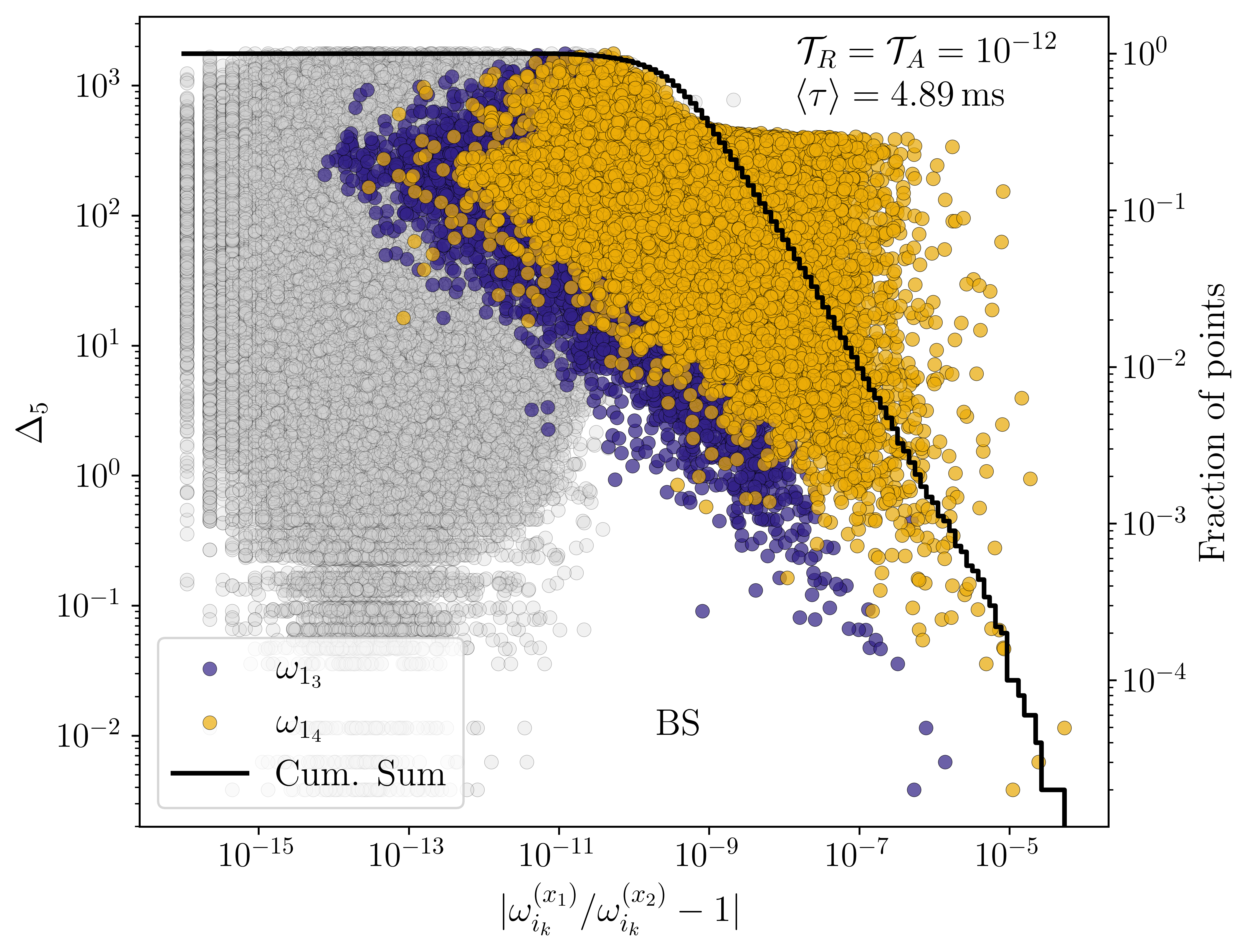}
    }
    \caption{Relative difference between final points from two distinct initial conditions versus the pentagon Gram determinant $\Delta_5$. The blue and yellow points denote the relative difference in each phase space point of the graded functions of transcendental weight three and four associated with the pentagon integral, $\omega_{1_3}$ and $\omega_{1_4}$, while the gray points depict all the other graded functions. The thick black line displays the cumulative sum of relative differences of the graded function with worst precision for each kinematic point, including all 65 functions. The 50k phase-space values have been generated with \texttt{Phokhara}. The average time and absolute and relative tolerances are also shown.}
    \label{fig:difNN}
\end{figure}

To the best of our knowledge, there is no software capable of doing a large scale computation of these integrals in a small amount of time,\footnote{We have tried to use the program \texttt{DCT} \cite{Huang:2024qan}. Although it is fast, with an average run-time of $\thicksim6$ms, it gives no result for certain kinematic configurations.} and we leave a comparison with some of the capable but slower tools for future work. Instead, we provide an estimate of the error by calculating the same final point starting from two different initial conditions. The relative difference between the two final values, which should be identical, does not provide a rigorous definition of the uncertainty associated with the calculation, but rather an estimate, and a sanity check that the integrator does not suffer from numerical instabilities. We evolve the differential equations from two initial conditions, keeping the electron and pion mass fixed to their physical values, and set the Mandelstam variables at the arbitrary kinematic points,
\begin{align}
    \vec{x}_1 &= \left\{\frac{2601}{2500}, -\frac{21531256}{60641275}, \frac{32330399}{58933369}, \frac{101859785}{220131679}, -\frac{10194155}{69478379}\right\}\,, \\
    \vec{x}_2 &= \left\{14, -\frac{5}{2}, \frac{11}{5}, \frac{4}{5}, -\frac{11}{10}\right\}\,.
\end{align}

~

Furthermore, to check that the integrator can be used in Monte Carlo simulations, we obtain 50 thousand phase-space points with the Monte Carlo event generator \texttt{Phokhara}~\cite{Campanario:2019mjh}, using the realistic scenario for B-meson factories from \cite{Aliberti:2024fpq}. Using the BS algorithm, we compute the relative difference between the final points, evaluated starting from $\vec{x}_1$ or $\vec{x}_2$, and plot the cumulative distribution in Fig.~\ref{fig:difNN}, taking the relative precision value of the graded function with the worst error at each phase space point. We also plot the more complicated graded functions of transcendental weight three and four that involve a pentagon integral in Fig.~\ref{fig:difNN}, versus the pentagon Gram determinant $\Delta_5$. The plot reinforces the point made in Section~\ref{sec:numerror}: as the final desired kinematics lie closer to a singularity, in this case $\Delta_5~=~0$ present in $\omega_{1_3}=\mathrm{Coef}_{\epsilon}(\epsilon^3\Delta_5 I_{1,1,1,1,1}^{NN})$ and $\omega_{1_4}=\mathrm{Coef}_{\epsilon^2}(\epsilon^3\Delta_5 I_{1,1,1,1,1}^{NN})$, the numerical error increases. The decrease on numerical precision due to $\Delta_5$ has also been observed in similar two-loop processes \cite{Chicherin:2021dyp, Abreu:2023rco}.  

\subsection{Two-loop five-point integrals for $t\bar{t}j$}

As a final test, we implement two families from~\cite{Badger:2024fgb} in our integrator, particularly PB$_A$ and PB$_B$, for the process $p\bar{p} \to t\bar{t}j$. It involves six kinematic variables, and the two families are composed of 88 and 121 master integrals, respectively, all needed at five orders in the dimensional regulator. For the first integral family, PB$_A$, a canonical form (see Eq.~\eqref{eq:canonical}) is provided. This is different for PB$_B$, where instead an $\epsilon$-factorised form is provided with dependence on both logarithmic and one-forms, following Eq.~\eqref{eq:oneforms}. For each integral, we evolve each order of $\epsilon$ in the six kinematic variables from a boundary point to the provided benchmark value. Note that in a realistic implementation, there would be no need to evolve the centre-of-mass energy or the mass. This will not only speed up the integrator by $\thicksim33\%$, but also the precision will improve. Furthermore, we could skip the integration of the graded functions at transcendental weight one, as they can be expressed as logarithms. It might also be possible to express the weight two functions as one-fold integrations. Nevertheless, one needs to remember that both weight one and weight two are also needed for integrating weight three and four functions and are thus calculated at each step of the numerical integration. Consequently, it might be the case that the evolution is actually slower (albeit more precise) when weight one and two are calculated analytically. Finally, we expect the use of graded functions to reduce the overall complexity of the integration, due to the organisation of transcendental functions and integration kernels.

~

We report in Table~\ref{tab:timing2lttj} the timing achieved by our integrator, as well as the significant figures $\mathcal{R}$  defined as \begin{equation}
    \mathcal{R} = -\log_{10}\left(\left|\frac{\text{ThisWork}}{\texttt{DiffExp}}-1\right|\right)\,.
\end{equation} Note that this formula is only used when the values from \texttt{DiffExp} are non zero, while the absolute error is applied in the other cases. Furthermore, these values should be taken as a reference, but with caution, since we expect a spread on the precision similar to what we have observed in previous examples, varying with the kinematics of the final point.

\begin{table}[t]
    \centering
    \setlength{\tabcolsep}{8pt}
    \renewcommand{\arraystretch}{1.3}
    \begin{tabular}{c c  c c c c}
    \toprule
        & & $\mathcal{T}_A$, $\mathcal{T}_R$ & $\mathcal{R}$ & $\langle \tau \rangle$ [s] & \texttt{DiffExp} $\langle \tau \rangle$ [s] \\ 
        \midrule
        \multirow{2}{*}{PB$_A$} 
        & \texttt{double} & $10^{-12},\ 10^{-12}$ & 10 & 0.0881 & 580.85 \\
        & \texttt{quad}   & $10^{-28},\ 10^{-28}$ & 27 & 51.588 & 795.516 \\
        \midrule
        \multirow{2}{*}{PB$_B$}
        & \texttt{double} & $10^{-12},\ 10^{-12}$  & 10 & 0.100  & 555.438 \\ 
        & \texttt{quad}   & $10^{-28},\ 10^{-28}$ & 27 & 89.088 & 826.219 \\
        \bottomrule
    \end{tabular}
    \caption{Performance summary for the evolution from a boundary point to a benchmark kinematic configuration in two-loop families, using double and quadruple precision. Integration was performed with the BS algorithm. The average integrator time in double precision corresponds to the mean over 1000 runs. Shown are the absolute and relative tolerances $(\mathcal{T}_A, \mathcal{T}_R)$, the number of correct digits $\mathcal{R}$, and the run-time of our integrator and the \texttt{DiffExp} package.}
    \label{tab:timing2lttj}
\end{table}

\section{Avenues for improvement} \label{sec:futurework}

We dedicate this section to discussing how precision and speed could be improved. The possibilities are many, but we concentrate our discussion around three points: the usage of a grid of boundary values, a different algorithm for choosing a path, and comments on general software improvements. 

The choice of initial conditions determines the amount of steps required to reach the final point, and with fewer steps, the run times are reduced. This motivates generating a grid of possible initial points and starting from the one that minimises the distance to travel. This could be exploited further: with a careful analysis of the singularities, we could generate a relatively small grid of initial conditions in different regions such that any final point is connected to an initial value without the need of an analytic continuation.
Together with the newly developed method for splitting real and imaginary components of Feynman integrals~\cite{Jones:2025jzc}, it would allow us to work with reals instead of complex numbers, which would, most likely, make a great improvement to performance. A grid of initial conditions could also be used near singularities to minimise numerical cancellations and ensure a faster and more precise solution. Together with extrapolation, higher precision types or combined with precomputed series solutions, we can ensure the precision needs are met. 
Finally, extensive use of graded functions is also beneficial. The spurious singularities are analytically dealt with, and the analytic structure of the amplitude shows nice properties, which allows us to evaluate only the integrals we need, instead of the full set.

With respect to finding a path, it would be interesting to explore the algorithm described in \cite{Chicherin:2021dyp}, or include a strategy that tries to minimise the integrated variation of the field along the path. We also expect an overall simplification and reduction in run times when one focuses in the physical region. 

There is clear room for improvement with regard to the integrator itself, both in the software and in the numerical techniques used. The current integrator can be further optimised, the implementation of quadruple precision can be enhanced, and other libraries for high-precision types should be explored. Furthermore, the use of quadruple precision only in targeted parts of the code, such as the computation of coefficients $a,b$ described in Section \ref{sec:reducetime}, would most likely improve precision with comparatively little computational overhead. In addition, there is the possibility of exploring other numerical algorithms. As seen in Section~\ref{sec:ResultsMM}, the choice between algorithms can significantly vary CPU times, an effect that we have observed to be enhanced for more complicated differential equations. It is not unreasonable to think that significant improvement can be made by developing a custom integrator, or using other libraries --- such as the one used in \cite{Czakon:2008zk}---, and exploring alternative numerical algorithms. For example, the fast convergence of spectral methods, already used extensively in numerical relativity \cite{Grandclement:2007sb}, makes it a promising candidate to explore. For large basis of master integrals, GPU computing might be a good option, which can be easily exploited in the numerical approach. Finally, it would be interesting to compare precision and efficiency with other numerical approaches, such as the ones relying on one-fold integration~\cite{Caron-Huot:2014lda}.

\section{Conclusions}

In this work, we elaborated on the numerical evaluation of Feynman integrals using the method of differential equations, focusing on the analytic structure of these integrals. Our approach exploits two representations of the differential equations: one in which the dimensional regulator $\epsilon$ factors out---yielding canonical differential equations---and another where the dependence on $\epsilon$ is strictly polynomial. In both representations, we accounted for treatment of algebraic functions. These formulations, respectively, rely on integration kernels expressed in terms of logarithmic forms and differential one-forms. Our framework allows for the computation of the transcendental functions that appear explicitly in physical scattering amplitudes, while avoiding spurious contributions that cancel at the level of the full amplitude.

We demonstrated the applicability of this method through numerical evaluations of representative examples: five-point one- and two-loop Feynman integrals involving up to nine independent complex kinematic scales, as well as the gauge-invariant pentabox piece for the amplitude of the process $e^+e^- \to \pi^+ \pi^- \gamma$ in scalar QED, evaluated at higher orders in the dimensional regulator. In the latter case, we illustrated the advantage of working with transcendental functions organised by weight grading, as this leads to more efficient and stable evaluations.

Our \texttt{C++} implementation of the numerical solution of differential equations incorporates three algorithms (specifically \texttt{runge\_kutta\_cash\_karp54}, \texttt{runge\_kutta\_dopri5} and  \texttt{bulirsch\_stoer}), which we systematically compared in terms of performance and stability. We recognised BS as the fastest and most precise algorithm. In addition, we overcame the challenges inherited by analytic continuation across different kinematic regions, by introducing a novel strategy that evolves one kinematic variable at a time, ensuring robustness even in the presence of many scales.

~

Building on our implementation, we identify promising directions for future work:

\begin{enumerate}
\item Our long-term goal is to embed this \texttt{C++} framework within Monte Carlo tools such as {\tt Phokhara}, thereby enabling fast and precise evaluation of multi-loop corrections directly in event generation. In particular, we aim to extend {\tt Phokhara}’s current next-to-leading order (NLO) capabilities to next-to-next-to-leading order (NNLO), which is crucial for precision studies of hadronic production in low-energy $e^+e^-$ collisions.

\item Depending on the physical observable and required accuracy, a better understanding of the necessary numerical precision is essential. While we have briefly explored transitions from double to quadruple precision, a more systematic study is needed to balance computational cost and numerical accuracy for different physical configurations. Custom improvements can be made for each system of differential equations, some of the possibilities involving grids of initial conditions, extrapolation, or reweighing of the kinematic scales. In the cases where on-the-fly evaluation is not possible, the integrator can be easily used to generate dense grids. 

\item Although our current implementation focuses on systems free of poles in $\epsilon$, this is not always feasible in more general settings. In such cases, it becomes necessary to find suitable rotations of the basis of master integrals that reveal a canonical form. Our current setup provides a solid framework for implementing (and testing) strategies to automate this reorganisation systematically.

\end{enumerate}

All these directions are natural extensions of the method proposed in this work and can be pursued within the unified framework we have developed. Requiring only the differential equation and an initial condition  as input, the integrator is capable of accommodating systems that are polynomial in $\epsilon$, 
delivering numerical values for the relevant Feynman integrals with promising runtimes for the processes studied. 
As a result of its flexibility, the integrator can easily include differential equations beyond those considered in this work. 
With a careful analysis of the precision and further optimisation, we envision the integrator will be a key ingredient towards implementing NNLO calculations in Monte Carlo event generators --- whether through fast on-the-fly evaluations or the efficient generation of interpolating grids.

\acknowledgments

We wish to thanks Daniel Gerardo Melo~Porras for cross-checking parts of the results, and Sophie~Kollatzsch, Yannick~Ulrich, Francesco~Pio~Ucci and Marco~Ghilardi for sharing preliminary results of the amplitude $e^+e^- \to \pi^+ \pi^- \gamma$ in scalar QED.
We would like to thank Philipp~Rendler and Antonela~Matijašić
for enlightening discussions in the workshop ``Scattering Amplitudes at Liverpool'', and Micha\l{} Czakon, Renato Prisco, Jonathan Ronca and Francesco Tramontano for comments on the manuscript. This work was supported by the Leverhulme Trust, LIP-2021-014.

\bibliographystyle{JHEP}
\bibliography{biblio.bib}

\end{document}